\documentclass[preprint]{aastex}
\usepackage{emulateapj5}
\begin{document}

\slugcomment{}

\title{A Near-Infrared Wide-Field Proper Motion Search for Brown Dwarfs}
\author{Joannah L. Hinz, Donald W. McCarthy, Jr.}
\affil{Steward Observatory, University of Arizona, Tucson, AZ  85721}
\email{jhinz@as.arizona.edu, mccarthy@as.arizona.edu}

\author{Doug A. Simons}
\affil{Gemini Observatory, Northern Operations Center, 670 N. A'ohoku Place, Hilo, HI  96720}
\email{dsimons@gemini.edu}

\author{Todd J. Henry}
\affil{Department of Physics and Astronomy, Georgia State University, Atlanta, GA  30303}
\email{thenry@chara.gsu.edu}

\author{J. Davy Kirkpatrick}
\affil{IPAC, Caltech, MS 100-22, Pasadena, CA  91125}
\email{davy@ipac.caltech.edu}

\and

\author {Patrick C. McGuire}
\affil{University of Bielefeld, Bielefeld, Germany D-33615}
\email{mcguire@physik.uni-bielefeld.de}

\begin{abstract}
A common proper motion survey of M dwarf stars within 8\,pc of the Sun reveals
no new stellar or brown dwarf companions at wide separations 
($\sim$\,100-1400\,AU). This survey tests whether the brown dwarf ``desert'' 
extends to large separations around M dwarf stars and further explores the
census of the solar neighborhood.  The sample includes 66 stars 
north of $-30\degr$ and within 8\,pc of the Sun.  Existing first epoch images 
are compared to new $J$-band images of the same fields an average of 7 years 
later to reveal proper motion companions within a $\sim$\,4 arcminute 
radius of the primary star.  No new companions are detected to a $J$-band 
limiting magnitude of $\sim$\,16.5, corresponding to a companion mass of 
$\sim$\,40 Jupiter masses for an assumed age of 5\,Gyr at the mean distance of 
the objects in the survey, 5.8\,pc.

\end{abstract}


\section{Introduction}
Although the sub-stellar initial mass function (IMF) has been studied in
a range of environments such as star-forming clusters 
(Luhman et al. 1998; Luhman 2000; Najita et al. 2000), 
young open clusters (Bouvier et al. 1998; Barrado y Navascues
et al. 2001) and the field (Reid et al. 1999), the IMF of low-mass companions 
is not well understood, especially at ``wide'' ($>$\,100\,AU) separations.
Radial velocity searches around solar-type main sequence stars 
(e.g., Mayor \& Queloz 1995; Marcy \& Butler 1996) have produced few 
confirmed brown dwarfs at separations $<$\,3\,AU.  Fewer than 0.5\% of
their sample have brown dwarf companions at those separations.  
A coronagraphic search for companions in the range 40-100\,AU 
(Oppenheimer et al. 2001) produced only one brown dwarf, GJ 229B 
(Nakajima et al. 1995), well below the 17-30\% multiplicity observed
for all stars (Reid \& Gizis 1997).  Other types of surveys, such
as high spatial resolution space-based observations (Lowrance et al. 1999;
Lowrance et al. 2000) and ground-based adaptive optics (Els et al. 2001),
have also resulted in discoveries of low-mass
stellar and sub-stellar companions.  However, the frequency of stellar
and sub-stellar companions at close separations remains distinctly 
different, resulting in the idea that there is a ``brown dwarf desert''.

To date there has been only one systematic search for brown dwarf
companions at wide separations and with a volume-limited sample (Simons
et al. 1996; hereafter, SHK).  This was mainly a color-based search around 
M dwarfs
within 8\,pc of the Sun and did not turn up any new brown dwarfs, although, 
given the surprisingly blue colors of GJ 229B, cool brown dwarfs with 
intermediate $J$-$K$ colors may have been overlooked in the survey. 

Proper motion searches for companions have been used for many years to identify
low-mass objects (e.g., van Biesbroeck 1961) and offer a less biased 
way of finding low-mass companions than color-based surveys.  Therefore, we 
have conducted the planned second epoch survey of the SHK sample, in order to 
identify 
low-mass companions to M dwarfs at wide separations out to over 1000\,AU.
The choice of M dwarf primaries is significant:  Reid \& Gizis (1997) and 
Reid et al. (1999) show that the distribution
of mass ratios for a sample of 80\% M dwarfs has a peak at $q=$ 0.95, where
$q$ is the ratio of the secondary mass to the primary mass.  They conclude
that their sample shows a distinct bias towards approximately equal-mass 
systems and that the mass function for stellar companions is different from the
IMF of field stars.  If these conclusions extend to brown dwarf masses, 
M primaries may harbor more sub-stellar companions than
other stellar types.  On the other hand, Reipurth \& Clarke (2001) suggest
that brown dwarfs have been ejected by dynamical interactions during
the star formation process and cannot accrete enough mass to become stars.
In this case M dwarf primaries may not be accompanied by such companions
except in a multiple M dwarf systems with a correspondingly large
gravitational potential.

Thus, our proper motion search around one
spectral class of primaries fills a unique niche in the search for low-mass
stellar and brown dwarf companions.  We describe the data acquisition and
reduction in $\S$\,2 and discuss the results of the survey in $\S$\,3.

\section{Observations and Data Reduction}
\subsection{Sample Selection}
Our sample is identical to Henry's (1991) list of M dwarfs within 8\,pc
of the Sun. 		  
The M dwarfs were chosen initially from the Second Catalog of Nearby Stars 
(Gliese 1969) and its updates (Gliese \& Jahreiss 1979) along with other 
additions from more recent literature (e.g., LHS 292; see SHK for more
details).  The sample consists of 75 M 
dwarf primaries with M$_V$\,$\geq$\,8.0 mag, trigonometric parallaxes 
$\geq$\,0$\farcs$125 and declinations north of $-30\degr$.  
SHK observed 66 of these systems, discarding three due to confusion toward
the galactic plane (GJ 701, GJ 729, and GJ 752), 
leaving a total of 63 systems.  

Of the original 75 systems, we observed 
74 targets.  Table 1 contains a list of the targets, their proper motions, 
the dates of observation and those of SHK, and other relevant parameters.
The median distance of the M dwarf targets is 5.8\,pc, corresponding to a
median search radius of 1480\,AU in the present survey.  Over the complete
distance range, the search radius varies from 800-2100\,AU.
Figure 1 shows the distribution of total proper motions of our objects 
between the first and second epoch
observations.  Due to discoveries of new objects and to the measurement of
more accurate parallaxes, the SHK list is no longer a complete
volume-limited sample.  Table 1 has five objects from the original sample
whose updated parallaxes move them beyond the 8\,pc limit (GJ 185, GJ 623,
GJ 686, GJ 1230, and GJ 884) and separately lists four objects whose 
redetermined parallaxes or recent discoveries (e.g., G 180-060; Ducourant et 
al. 1998) place them within the survey criteria. 

\subsection{Imaging}
First epoch images (SHK) were obtained at the University of Hawaii's 24-inch
telescope between 1991 August and 1992 August with a facility 256 x 256 
NICMOS camera in both the $J$ and $K'$ bands with a scale of 
2$\farcs$0\,pixel$^{-1}$.  Exposure times were typically 1 hour, and images 
were processed using conventional techniques.  A custom program searched
for point sources above a 3\,$\sigma$ detection level and performed photometry
on all sources using a 10$\arcsec$ aperture.

Between 1998 April and 2000 December,
$J$-band images of the same fields were taken with the PISCES camera (McCarthy
et al. 2001) at the Bok 2.3\,m telescope on Kitt Peak.  PISCES has an 
8$\farcm$5 diameter field-of-view and a 0$\farcs$5\,pixel$^{-1}$ plate scale
at this telescope.  Nine 30 second exposures were obtained, centered on a
program M dwarf, with a 10$\arcsec$ dither between each exposure.  All images
were corrected for quadrant cross-talk effects known 
to be present in HAWAII arrays (McCarthy et al. 2001).
The images were then dark-subtracted, flat-fielded, masked for hot pixels, 
corrected for geometric distortion, and combined with standard IRAF tasks.  
The flat-field was produced through a median combination of the dithered 
science frames.  The flat-field is predictably poor in the region of the
bright M star, which was allowed to saturate the detector.  However, the
flat-fielding does not affect the astrometry of the field, changing the 
position of the M dwarf by less than half a pixel in multiple test cases. Also,
because accurate photometry of the objects has already been carried out by
SHK, the resulting flat-field is adequate for this survey.

Figure 2 shows a sample fully reduced field (GJ 752) along with the
identical field observed by SHK.  The field-of-view sizes are almost
perfectly matched, except that PISCES has a circular inscribed field.
The SHK images show an internal reflection due to the optics in the camera 
to the lower right of the M dwarf primary that is not in the second
epoch set.  The SHK survey has a limiting
magnitude of m$_J$\,$\sim$\,16.5 and is sensitive to companions down to 
40 M$_{Jupiter}$ assuming an age of 5\,Gyr at the mean distance of the survey,
5.8\,pc, based on the models of
Burrows et al. (1997).  We use 5\,Gyr, following the findings of Henry (1991)
on this sample; however, because age dating M dwarfs is difficult, 5\,Gyr may
not be an accurate average age.  If the M dwarfs are instead 1\,Gyr old, 
the survey is sensitive to 16 M$_{Jupiter}$.  The sensitivity 
of the PISCES images matches or exceeds that of the SHK images in all cases 
with m$_J$\,$\sim$\,17.0.

Each second epoch image is compared to its matching first epoch image 
using an IRAF script originally designed to identify supernovae in nearby 
galaxies (Van Dyk et al. 2000).  Using input coordinates of identical
objects in the two frames, the script matches the pixel scales of the two
cameras, accounts for any differences in geometric distortion, matches the
point spread functions for the two images, and subtracts them.  Typically,
10-15 background stars from each M dwarf field were used as input references
for the program.  The total number of background stars ranges from 50
to over 1000 sources for the crowded fields near the galactic plane.
The mean total proper motion of the M dwarfs between epochs 
is $8\farcs8$ ($\sim$\,18 pixels on the PISCES camera).  Moving objects are
revealed by adjacent positive and negative images.  Companion objects would 
have proper motion vectors identical to the M dwarf primaries which have
been accurately measured by Hipparcos (Perryman et al. 1997).  
Objects with such large proper motions can easily be detected by visually
examining the subtracted images.

Figure 3 shows the detection of the known low-mass companion van Biesbroeck
10 (VB 10, GJ 752B).  Other known wide companions were also detected as 
indicated with an asterisk
in Table 1.  These results demonstrate the reliability of the subtraction 
method.  GJ 570A, known to have a T dwarf companion at a separation of 
258$\farcs$3 (Burgasser et al. 2000), was not included in the sample because
the primary is a K4 dwarf.

\section{Results}
No new low-mass stellar or brown dwarf companions were detected in this
8\,pc sample of M dwarfs.  The same conclusion was reached by SHK from
a $J$-$K'$ color search.  However, as many as nine wide 
(3600\,$\ge$\,$\Delta$\,$\ge$\,120\,AU) companions have recently 
been detected around nearby (9.6-39\,pc) stars using the Two Micron 
All-Sky Survey (2MASS) database (Kirkpatrick et al. 2000, 2001; 
Burgasser et al. 2000; Wilson et al. 2001). Seven of these new objects 
are common proper motion companions. In general, it is difficult to 
compare the 2MASS results with the present survey because the
2MASS primary stars have uncertain ages and generally higher luminosities.
Based on initial 2MASS results, Gizis et al. (2001) estimate that 
$\sim$\,1\% of primaries with masses 0.6-1.5\,M$_{\odot}$ (M$_V$\,$<$\,9.5) 
have 
wide ($\ge$\,1000\,AU) L dwarf companions and that the frequency of all 
wide brown dwarf companions is 5-13 times greater.  Extending this 
analysis to our M dwarf sample of 63 objects with masses between 
0.08-0.6\,M$_{\odot}$, we would expect to detect between 3 (5\% of our sample) 
and 9 (13\%) brown dwarf companions.  The apparent difference between our 
results and Gizis et al. (2001) might be resolved if the frequency of brown
dwarf companions depends strongly on primary mass and orbital separation.
This possibility could be tested either by systematic data mining of the
2MASS survey or by extending the present PISCES survey to other spectral
types.

\acknowledgments
J. L. H. thanks Chien Peng, Rose Finn, and Roelof de Jong for valuable
discussions concerning near-infrared data reduction and Michael Meyer for
comments on early drafts of this paper.  The HAWAII detector and electronic
controller for PISCES were purchased by the Air Force Office of Scientific 
Research through grant F49620-96-1-0285.  The PISCES camera was also briefly
supported by the National Science Foundation through grant NSF 96-23788.
D. A. S. was supported by the Gemini Observatory, which is
operated by the Association of Universities for Research in Astronomy,
Inc., under a cooperative agreement with the NSF on behalf of the Gemini
partnership: the National Science Foundation (United States), the
Particle Physics and Astronomy Research Council (United Kingdom), the
National Research Council (Canada), CONICYT (Chile), the Australian
Research Council (Australia), CNPq (Brazil) and CONICET (Argentina).

\clearpage
\begin{deluxetable}{cccccccccccccccc}
\tabletypesize{\tiny}
\tablecaption{Parameter list for target program stars.}
\tablewidth{400pt}
\tablehead{
\colhead{Primary Name} & \colhead{Components}  & 
\colhead{Trig. Parallax} & \colhead{Proper Motion} & 
\colhead{1$^{st}$ Epoch}  & \colhead{2$^{nd}$ Epoch} & 
\colhead{M$_V$} & \colhead{Sp. Type}}
\startdata
GJ 1002 &       &.2128$\pm$.0033 & 2.041 & ---      & Jan 1999 & 15.4  & M5.5\\
GJ 1005 & AB    &.1919$\pm$.0172 & 0.863 & Jan 1992 & Jan 1999 & 12.9  & M4.0\\
GJ 15   & AB*    &.2802$\pm$.0011 & 2.912 & Aug 1991 & Jan 1999 & 10.3  & M1.5\\
GJ 2005 & ABCD  &.1328$\pm$.0091 & 0.614 & ---      & Jan 1999 & 15.4  & M5.5\\
GJ 54.1 &       &.2690$\pm$.0076 & 1.345 & ---      & Jan 1999 & 13.7  & M4.5\\
GJ 65   & AB    &.3807$\pm$.0043 & 3.368 & Jan 1992 & Jan 1999 & 15.4  & M5.5\\
GJ 83.1 &       &.2238$\pm$.0029 & 2.907 & Aug 1991 & Jan 1999 & 14.0  & M4.5\\
GJ 109  &       &.1324$\pm$.0025 & 0.923 & Jan 1992 & Jan 1999 & 11.2  & M3.0\\
GJ 185  & AB    &.1203$\pm$.0017 & 0.308 & Feb 1992 & Jan 1999 & 8.9   & K7.0\\
GJ 205  &       &.1757$\pm$.0012 & 2.235 & Jan 1992 & Dec 1998 & 9.1   & M1.5\\
GJ 213$^{\dagger}$  &       &.1728$\pm$.0039 & 2.571 & Oct 1991 & Dec 1998 & 12.7  & M4.0\\
LHS 1805 &      &.1322$\pm$.0029 & 0.831 & ---      & Dec 1998 & 12.3  & M3.5\\
G 099-049 &       &.1863$\pm$.0062 & 0.241 & ---      & Dec 1998 & 12.7  & M3.5\\
GJ 229  & AB    &.1732$\pm$.0011 & 0.737 & Mar 1992 & Dec 1998 & 9.3   & M1.0\\
GJ 234  & AB    &.2429$\pm$.0026 & 0.997 & Feb 1992 & Dec 1998 & 13.0  & M4.5\\
GJ 251  &       &.1813$\pm$.0019 & 0.851 & Jan 1992 & Dec 1998 & 11.2  & M3.0\\
GJ 1093 &       &.1289$\pm$.0035 & 1.225 & ---      & Jan 1999 & 15.4  & M5.0\\
GJ 268  & AB    &.1572$\pm$.0033 & 1.052 & Jan 1992 & Jan 1999 & 12.5  & M4.5\\
GJ 273  &       &.2633$\pm$.0014 & 3.761 & Jan 1992 & Jan 1999 & 12.0  & M3.5\\
GJ 285  &       &.1686$\pm$.0027 & 0.604 & Mar 1992 & Jan 1999 & 12.3  & M4.0\\
GJ 299  &       &.1480$\pm$.0026 & 5.211 & Mar 1992 & Jan 1999 & 13.7  & M4.0\\
GJ 300  &       &.1700$\pm$.0102 & 0.707 & Jan 1992 & ---      & 14.2  & M3.5\\
GJ 1111 &       &.2758$\pm$.0030 & 1.29  & Jan 1992 & Jan 1999 & 17.0  & M6.5\\
GJ 1116 & AB    &.1913$\pm$.0025 & 0.874 & Jan 1992 & Jan 1999 & 15.5  & M5.5\\
GJ 338  & AB*    &.1616$\pm$.0052 & 1.662 & Jan 1992 & Jan 1999 & 8.7   & M0.0\\
GJ 380  &       &.2052$\pm$.0008 & 1.454 & ---      & Jan 1999 & 8.2   & K7.0\\
GJ 388  &       &.2039$\pm$.0028 & 0.506 & Feb 1992 & Jan 1999 & 11.0  & M3.0\\
GJ 393  &       &.1383$\pm$.0021 & 0.949 & Mar 1992 & Jan 1999 & 10.3  & M2.0\\
LHS 292 &       &.2210$\pm$.0036 & 1.644 & Feb 1992 & Jan 1999 & 17.3  & M6.5\\
GJ 402  &       &.1775$\pm$.0230 & 1.15  & Mar 1992 & Jan 1999 & 12.9  & M4.0\\
GJ 406  &       &.4183$\pm$.0025 & 4.696 & Mar 1992 & Jan 1999 & 16.6  & M6.0\\
GJ 408  &       &.1510$\pm$.0016 & 0.465 & Mar 1992 & Jan 1999 & 10.9  & M2.5\\
GJ 411  &       &.3925$\pm$.0009 & 4.807 & Jan 1992 & Jan 1999 & 10.5  & M2.0\\
GJ 412  & AB*    &.2069$\pm$.0012 & 4.528 & Mar 1992 & Jan 1999 & 10.3  & M1.0\\
GJ 445  &       &.1855$\pm$.0014 & 0.863 & Feb 1992 & Jan 1999 & 12.1  & M3.5\\
GJ 447  &       &.2996$\pm$.0022 & 1.348 & Mar 1992 & Jan 1999 & 13.4  & M4.0\\
GJ 1156 &       &.1529$\pm$.0030 & 1.301 & ---      & Jan 1999 & 14.7  & M5.0\\
GJ 473$^{\dagger}$  & AB    &.2322$\pm$.0043 & 1.811 & Mar 1992 & Jan 2000 & 14.3  & M5.5\\
GJ 514  &       &.1311$\pm$.0013 & 1.552 & Jan 1992 & Jan 2000 & 9.6   & M1.0\\
GJ 526  &       &.1841$\pm$.0013 & 2.325 & Mar 1992 & Jan 2000 & 9.8   & M1.5\\
GJ 555$^{\dagger}$  &       &.1635$\pm$.0028 & 0.69  & Mar 1992 & Jul 1998 & 12.4  & M3.5\\
LHS 3003$^{\dagger}$ &      &.1610$\pm$.0060 & 0.965 & ---      & Jul 1998 & 18.1  & M7.0\\
GJ 581$^{\dagger}$  &       &.1595$\pm$.0023 & 1.224 & Mar 1992 & Jul 1998 & 11.6  & M2.5\\
GJ 623$^{\dagger}$  & AB    &.1243$\pm$.0012 & 1.231 & Aug 1991 & Jul 1998 & 10.7  & M2.5\\
GJ 625$^{\dagger}$  &       &.1519$\pm$.0011 & 0.42  & Mar 1992 & May 1998 & 11.3  & M1.5\\
GJ 628  &       &.2345$\pm$.0018 & 1.175 & Mar 1992 & Jul 2000 & 12.0  & M3.0\\
GJ 644$^{\dagger}$  & ABCD+643* &.1539$\pm$.0026 & 1.183 & Aug 1992 & May 1998 & 10.7 & M2.5\\
G 203-047 & AB  &.1378$\pm$.0090 & 0.428 & ---      & May 1998 & 12.5  & M3.5\\
GJ 661$^{\dagger}$  & AB    &.1595$\pm$.0031 & 1.582 & Aug 1991 & May 1998 & 11.0 & M3.0\\
GJ 673  &       &.1295$\pm$.0010 & 1.315 & ---      & May 1998 & 8.1   & K7.0\\
GJ 686  &       &.1230$\pm$.0016 & 1.361 & ---      & May 1998 & 10.1  & M0.0\\
GJ 687  &       &.2209$\pm$.0009 & 1.304 & ---      & May 1998 & 10.9  & M3.0\\
GJ 699$^{\dagger}$  &       &.5493$\pm$.0016 & 10.31 & Aug 1991 & May 1998 & 13.2  & M4.0\\
GJ 701$^{\dagger}$  &       &.1283$\pm$.0014 & 0.644 & Aug 1992 & May 1998 & 9.9   & M0.0\\
GJ 1224 &       &.1327$\pm$.0037 & 0.664 & ---      & May 1998 & 14.3  & M4.5\\
LHS 3376 &      &.1373$\pm$.0053 & 0.623 & ---      & Jul 1998 & 14.1  & M4.5\\
GJ 1230 & ABC   &.1209$\pm$.0072 & 0.501 & Aug 1991 & May 1998 & 12.8  & M4.5\\
GJ 725  & AB*    &.2802$\pm$.0026 & 2.273 & Oct 1991 & May 1998 & 11.1  & M3.0\\
GJ 729$^{\dagger}$  &       &.3365$\pm$.0018 & 0.72  & Aug 1992 & May 1998 & 13.6  & M3.5\\
GJ 752  & AB*    &.1703$\pm$.0014 & 1.466 & Aug 1992 & May 1998 & 10.3  & M3.0\\
GJ 1245 & AB*C   &.2120$\pm$.0043 & 0.731 & Aug 1991 & Jul 1998 & 15.0  & M5.5\\
GJ 809  &       &.1420$\pm$.0008 & 0.772 & Oct 1991 & Nov 2000 & 9.3   & M0.0\\
GJ 829  & AB    &.1483$\pm$.0019 & 1.058 & Aug 1991 & Nov 2000 & 11.2  & M3.5\\
GJ 831  & ABC   &.1256$\pm$.0045 & 1.194 & Aug 1992 & Nov 2000 & 12.6  & M4.5\\
LHS 3799 &      &.1341$\pm$.0056 & 0.778 & ---      & Nov 2000 & 13.9  & M4.5\\
GJ 860  & AB    &.2495$\pm$.0030 & 0.943 & Oct 1991 & Nov 2000 & 11.6  & M3.0\\
GJ 866  & ABC   &.2943$\pm$.0035 & 3.254 & Aug 1992 & Nov 2000 & 14.4  & M5.0\\
GJ 873  & AB*    &.1981$\pm$.0021 & 0.901 & Aug 1991 & Nov 2000 & 11.6  & M3.5\\
GJ 876  & AB    &.2127$\pm$.0021 & 1.143 & Aug 1992 & Nov 2000 & 11.8  & M3.5\\
GJ 880  &       &.1453$\pm$.0012 & 1.071 & Aug 1992 & Nov 2000 & 9.5   & M1.5\\
GJ 884  &       &.1228$\pm$.0009 & 0.911 & ---      & Nov 2000 & 8.3   & K5.0\\
GJ 896  & ABCD  &.1601$\pm$.0028 & 0.56  & Aug 1992 & Nov 2000 & 11.3  & M3.5\\
GJ 1286 &       &.1386$\pm$.0035 & 1.157 & Aug 1992 & Nov 2000 & 15.4  & M5.5\\
GJ 905$^{\dagger}$  &       &.3156$\pm$.0016 & 1.617 & Aug 1991 & Nov 2000 & 14.8  & M5.5\\
GJ 908  &       &.1675$\pm$.0015 & 1.37  & Aug 1992 & Nov 2000 & 10.1  & M1.0\\
\cutinhead{M Dwarfs That Would Now Meet Survey Criteria}
GJ 382  &       &.1273$\pm$.0015 & 0.287 & ---      & ---      & 9.3  & M1.5\\
G 180-060 &     &.1560$\pm$.0040 & ---   & ---      & ---      & 14.8 & M5.0\\
GJ 793  &       &.1251$\pm$.0011 & 0.526 & ---      & Jul 1998 & 10.4 & M2.5\\
LP 816-060 &    &.1822$\pm$.0037 & 0.308 & ---      & Nov 2000 & 12.7 & M
\tablecomments{
Column header explanations:  1. Primary M star designation, where a $\dagger$ 
symbol denotes that the primary was not centered in the second epoch
image, so that the search radius is not exactly 4$\farcm$25, 2. Known 
companions, marked with an asterisk if detected in the proper motion search,
 3. Trigonometric parallax ($\arcsec$) from SHK or from
Perryman et al. (1997), 4. Proper motion
of primary star ($\arcsec$\,year$^{-1}$), 5. Date of first epoch observation, 
6. Date of second epoch observation, 7. Absolute $V$-band magnitude of the 
primary star, 8. Spectral type of the
primary star.}
\enddata
\end{deluxetable}

\clearpage
\figcaption[]{Histogram illustrating the distribution of total proper motions 
between epochs ($\arcsec$) of the M dwarfs observed in this survey.
The median total proper motion value is 8$\farcs$8, indicated above.}

\figcaption[]{Sample $J$-band images of GJ 752 from the first ({\it left}) and
second ({\it right}) epochs.  North is up and east is to the left.
The field-of-view is 8$\farcm$5 on a side,
corresponding to a radial separation of 1500\,AU from the primary star.  
The first epoch image (SHK) was obtained in 1992 and
has a pixel scale of 2$\farcs$1\,pixel$^{-1}$.  The second epoch image taken
with the PISCES camera has a pixel scale of 0$\farcs$5\,pixel$^{-1}$.
Limiting $J$-band magnitudes are $\sim$\,16.5 and 17, respectively.
The ring of flux in the SHK image to the lower right of the primary star is
an internal reflection in the optics of the infrared camera.}

\figcaption[]{Difference images obtained by subtracting the first and
second epoch images of GJ 752, shown in Figure 2.  
Results are shown with ({\it left}) and without
({\it right}) matching the PSFs of the two cameras before subtraction.
The adjacent positive and negative images 
of the primary star show the motion over $\sim$\,6 years.  The proper motion 
companion VB 10, m$_J$\,$\sim$\,9.90, is visible to the lower left of the 
primary star.}

\end{document}